\documentclass[aps,pre,twocolumn,showpacs,superscriptaddress,groupedaddress,floatfix]{revtex4-2}
\usepackage{amsmath}
\usepackage{amssymb}
\usepackage{amsfonts}
\usepackage{graphicx}
\usepackage{dcolumn}
\usepackage{sidecap}
\usepackage{parskip}
\usepackage{grffile}
\usepackage{color}
\usepackage[breaklinks]{hyperref}
\usepackage{hhline}
\usepackage{mathtools}
\usepackage{graphics}
\usepackage{multirow}
\usepackage{verbatim}
\usepackage{longtable}
\usepackage{rotating}
\usepackage{setspace}
\usepackage{epsfig}
\usepackage{subfigure}
\usepackage{epstopdf}
\usepackage[normalem]{ulem}
\makeatletter
\def\@eqnnum{{\normalsize \normalcolor (\theequation)}}
 \makeatother

\graphicspath{{./}{ER/main/}}

\begin{document}

\title{Smallworldness in Hypergraphs}
\author{Tanu Raghav$^{1}$, Stefano Boccaletti$^{1, 2, 3, 4}$, Sarika Jalan$^{1}$}
\affiliation{1.Complex Systems Lab, Indian Institute of Technology Indore - Simrol, Indore - 453552, India}
\affiliation{2.Universidad Rey Juan Carlos, Calle Tulip\'an s/n, 28933 M\'ostoles, Madrid, Spain}
\affiliation{3.CNR - Institute of Complex Systems, Via Madonna del Piano 10, I-50019 Sesto Fiorentino, Italy}
\affiliation{4.Moscow Institute of Physics and Technology, Dolgoprudny, Moscow Region, 141701, Russian Federation}

\date{\today}

\pacs{}

\begin{abstract}
Most real-world networks are endowed with the small-world property, by means of which the maximal distance between any two of their nodes scales logarithmically rather than linearly with their size. The evidence sparkled a wealth of studies trying to reveal possible mechanisms through which the pairwise interactions amongst the units of a network are structured in a way to determine such observed regularity.
Here we show that smallworldness occurs also when interactions are of higher order. Namely, by considering $Q$-uniform hypergraphs and a process through which connections can be randomly rewired with given probability $p$, we find that such systems may exhibit prominent clustering properties in connection with small average path lengths for a wide range of $p$ values, in analogy to the case of dyadic interactions. The nature of small-world transition remains the same at different orders $Q$ of the interactions, however, the increase in the hyperedge order reduces the range of rewiring probability for which smallworldness emerge.
\end{abstract}

\maketitle

In $1967$s Stanley Milgram diffused to the scientific community the outcome of his famous experiment involving $296$ individuals of the United States, where chains of letters were formed from a given "source" person to another given "target" person \cite{Milgram1967}. Milgram's results suggested that any target can be reached from any source by only a small number of steps, this way determining a small-world effect. It all started, as a curiosity, in $1929$ when Frigyes Karinthy wrote the short story "Chains" \cite{Karinthy1967}, which was followed by a series of mathematical studies by Pool and Kochen \cite{PoolKocher1978}. Studying the structure of mutual acquaintances across the world gave social networks a new standpoint, the concept of 'six degrees of separation' was introduced and later generalized to that of small-world networks.

In $1998$, Watts and Strogatz proposed a mechanism to generate such small-world networks:  starting from a regular lattice a fraction $p$ of links is randomly rewired. In this way, they succeeded to interpolate between a regular lattice ($p=0$) and a random graph ($p=1$) as the two limiting cases \cite{WattsStrogatz1998}, and found the presence of a wide range of $p$ values, where the resulting networks were simultaneously endowed with a high clustering (as high as that characterizing regular lattices) and at the same time an average path length comparable to that of the random graphs, which scales logarithmically with the number of vertices $N$ in the network. A substantial body of research exist in the literature suggesting that real-world networks such as power grids \cite{WattsStrogatz1998}, internet \cite{internet}, C. elegans \cite{WattsStrogatz1998}, air traffic \cite{airt}, polymers \cite{polymers}, brain \cite{brain} and metabolic pathways \cite{metabolic} display indeed such small-world phenomenon. Moreover, small-world networks are of interest because of their potential to explain the properties of the collective dynamics emerging on real-world networks \cite{new}.

In last two decades, many studies have adopted a network representation of various real-world systems, wherein interactions among elementary units were accounting for the underlying dynamics. However, when one adopts a network representation of a system, the assumption is made that the overall action of the entire system on each unitary component is  always factorizable into a combination of pairwise interactions.
The hypothesis may find justification when, for instance, the nature of the interaction is linear, but it is in fact very short in representing faithfully many other circumstances,  where instead higher-order interactions have to be taken into account for a proper description of the system.
The peculiar feature of higher order interactions is that a single hyperlink can connect more than just two nodes. Signature and thus the implications of such interactions have been observed in complex systems such as brain \cite{hbrain}, social \cite{hsocial}, ecological \cite{heco}, biological networks \cite{hbio}, evolutionary dynamics \cite{hevol} and protein interactions \cite{hpro}.

Network representations that embody higher order interactions are made via hypergraphs. A hypergraph $H=(V, E)$ is made of a node set $V=\{V_1, \cdots, V_N \}$ and of an hyperedge set $E=\{e_1, e_2, \cdots, e_M \}$, with $N$ and $M$ being, respectively, the number of vertices and hyperedges. $E$ is a multiset of $V$ where each subset is termed as a hyperedge. The number of nodes participating in a hyperedge \textit{$e_i$} is called cardinality (or order) of \textit{$e_i$}. An hyperedge of order $Q$ is a hyperlink connecting $Q$ vertices, thus standing for the group interaction of the corresponding $Q$ units of the network. Earlier studies on hypergraphs have obtained some success in modeling real-world systems such as brain \cite{hbrain}, protein interaction \cite{hpro}, social \cite{hsocial}, evolutionary dynamics \cite{hevol}, signaling pathways \cite{signal} etc.

\begin{figure}[t]
	\centering
	\includegraphics[width= 0.5\textwidth]{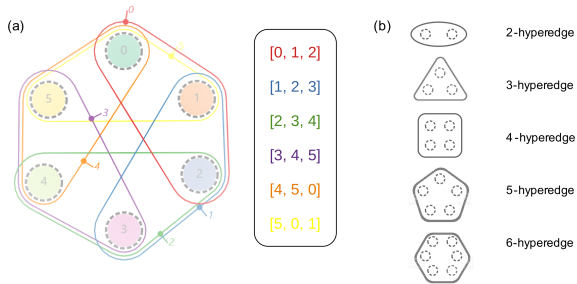}
    \vspace{-0.8cm}
	\caption{(color online) (a) Schematic representation of a $3-$ uniform hypergraph, arranged in a ring and formed by nearest neighbor hyperedges. The hyperedges are colored accordingly to the list appearing at the right of the panel. (b) Illustrations of hyperedges of different orders.}
	\label{sch1}
\end{figure}

In our work,  we will focus on uniform hypergraphs, where the same cardinality \textit{Q} characterize each hyperedge. \textit{Q}-uniform (or \textit{Q}-regular) hypergraphs have, therefore, all hyperedges that connect \textit{Q} nodes together. For instance, Fig.~\ref{sch1} depicts a $3$-uniform hypergraph. Note that a classic, pair-wise, network is just a 2-uniform hypergraph. Now, the hyperdegree $d_H(i)$ is the number of hyperedges incident on node $i$. In Figure.~\ref{sch1}, the hyperdegree of all nodes is $3$. In our work, we analyze  \textit{Q}-uniform hypergraphs for $Q=2, 3, 4, 5$ and $6$,  and compare their structural properties as, e.g., the clustering coefficient $C$ and the average shortest path length $L$. In particular, we will study how small-world states may emerge in such regular hypergraphs, and quantify the range for which they may occur.

\begin{figure}[b]
	\centering
	\includegraphics[width= 0.5\textwidth]{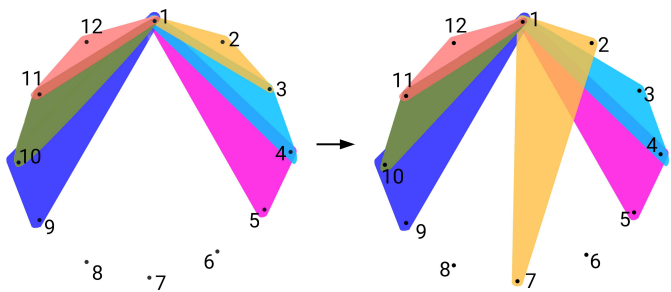}
    \vspace{-0.8cm}
	\caption{(Color online) Illustration of the random rewiring process. A $3$-uniform hypergraph with the hyperedges pertinent to node $1$ is shown on the left of the Figure: node $1$ participates in $2$ nearest neighbor hyperedges ($[1, 2, 3]$ and $[1, 12, 11]$), in $2$ next-nearest neighbor hyperedges $([1, 3, 4]$ and $[1, 11, 10])$ and in $2$ next-to-next nearest neighbor hyperedges $([1, 4, 5]$ and $ [1, 10, 9])$. In the right picture, node $7$ is selected randomly from the set of nodes that were not originally connected to node $1$ and, with probability $p$, the hyperedge $[1, 2, 3]$ is rewired to $[1, 2, 7]$.}
	\label{sch2}
\end{figure}

Let us start with a general framework for modeling regular hypergraphs (or hyper regular lattice) where each node is linked to a fixed number of nearest neighbor nodes by the same number of hyperedges. In order to construct a $Q$-uniform hypergraph, one then begins with $N$ vertices arranged in a ring fashion and label them (see Fig.~\ref{sch2} (left)). Starting from node $1$, one chooses the next  $Q-1$ nodes on both sides (clockwise and anticlockwise) to generate  $Q$ hyperedges, for example: hyperedges $[1, 2, 3]$ and $[1, 12, 11]$ in Fig.~\ref{sch2} (left). Further, in order to generate the next nearest neighbor hyperedges, one skips the nearest node and connects $Q-1$ nodes next to the nearest node, as in the case of the hyperedges $[1, 3, 4]$ and $[1, 11, 10]$. Similarly, for next-to-next nearest neighbor hyperedges, one skips two vicinal nodes of node $1$ and connect $Q-1$ nodes alongside them, thus obtaining hyperedges $[1, 4, 5]$ and $[1, 10, 9]$. The process is iterated in every node. As such, node $1$ has three originating hyperedges on both sides, in total $6$. Here, one further defines the degree of the $i_{th}$ node as the number of hyperedges originating in $i_{th}$ node. Note that the degree and hyperdegree of a node are two different measures, as the number of hyperedges originating from a node is different from the number of hyperedges incident on that node. For instance, for all nodes of Fig.~\ref{sch1}, the degree $k$ is equal to 2 while the hyperdegree $d_H$ is equal to 3.

Next, in order to cover the entire range between the two extremes (i.e., regular and random) configurations, hyperedges are randomized with a probability $p$. Inspired by the Watts-Strogatz algorithm \cite{WattsStrogatz1998}, we start by selecting a node $i$ (labeled as $1$ in Fig.~\ref{sch2}) and operate in a clockwise direction i.e.,  we choose the hyperedge that connects the $Q-1$ nearest neighbor nodes, say $[i, j, k]$ (edge $[1, 2, 3]$ in Fig.~\ref{sch2}). A node $z$ is then selected randomly (with uniform probability)  among all other nodes which are not connected to node $i$, and the farthest node from $i_{th}$ node in the hyperedge (i.e., the node labeled with the largest number in the hyperedge) is replaced by the new randomly chosen node. In other words,  the hyperedge $[i, j, k]$ is rewired with probability $p$ by replacing the node $k$ with a randomly selected node $z$ and becomes $[i, j, z]$. In Fig.~\ref{sch2} (right) it is shown how the hyperedge $[1, 2, 3]$ is rewired into $[1, 2, 7]$. The process is then continued for all the original hyperedges. This way the network structure can be calibrated from being completely regular (\textit{p}=$0$) to being purely random (\textit{p}=$1$), and analyzed for intermediate $p$, $0<\textit{p}<1$.

In order to properly describe the structural properties of the emerging hypergraphs, the characteristic path length $L(p)$ and the clustering coefficient $C(p)$ are measured. Here, $L(p)$ stands for the averaged minimum number of hyperedges required to reach a target vertex from a source vertex in the hypergraph. In Figure.~\ref{sch1}, the path length $L$ is $1$ from node $0$ to node $3$, because of the presence of the hyperedges $0$ and $1$ (or $4$ and $3$). The specific path (the sequence of hyperedges) corresponding to the shortest path length is also termed as hyperpath. $C(p)$ measures instead (on average) how likely it is that neighbors of a node are neighbors of each other. $C_{i}$ is defined as the ratio of the actual number of hyperedges between the neighbors of node $i$ to the possible number of edges between the neighbors. In all our trials, \textit{N} and \textit{k} are chosen such that the network remains connected for each value of $p$.
\begin{figure}[b]
	\centering
	\includegraphics[width= 0.5\textwidth]{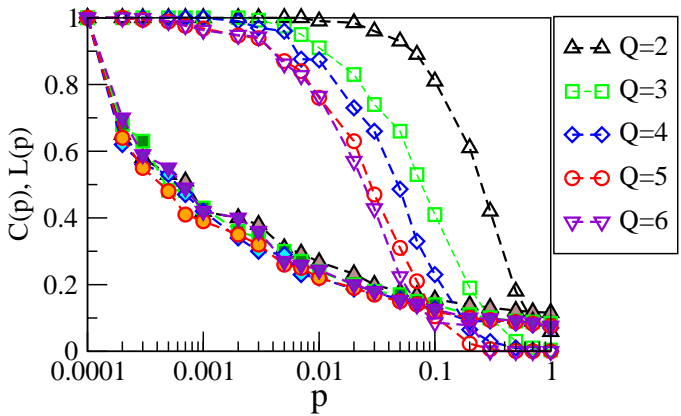}
	\caption{(Color online)  $C(p)$ (open symbol) and $L(p)$ (closed symbol) (see text for definitions) as a function of the rewiring probability $p$. Data are averaged over an ensemble of $20$ random realizations, and are further normalized by $C(0)$ and $L(0)$ values for hyper regular lattices. A logarithmic horizontal scale has been used and $N=500$ nodes and average degree $k=20$ hyperedges per node are taken for each hypergraph. The data is shown for different orders of uniform hypergraphs from $Q=2$ to $Q=6$, each designated with a different color. }
	\label{sw}
\end{figure}

First, we monitor the behavior of the clustering coefficient and average shortest path length as functions of the rewiring probability $p$. For pair-wise interactions, regular lattices ($p=0$) are known to be highly clustered and to display a linear scaling of $L$ as a function of $N$, while random networks ($p=1$) are poorly clustered and have $L$ that scales logarithmically with $N$.

\textit{Small-world phenomena:} Fig.~\ref{sw} reports $L(p)$ and $C(p)$ at different orders of uniform hypergraphs. In all cases, it is seen that a wide range of $p$ exists where $C($random$)<<C(p)\sim C($lattice$)$ and $L($random$)\sim L(p)<<L($lattice$)$. An increase in $p$ introduces long-range hyperedges causing a prompt fall in $L(p)$. Because of these long-range edges, vertices which are originally far apart from each other may be directly connected. The sudden drop in $L(p)$ not only reduces the distance between connected vertices but also reduces the distance between their neighbors, neighbors of neighbors, and so on. However, a rewired edge of a node does not cause an abrupt fall of $C(p)$ as observed in pair-wise networks. The result points to the fact that the small-world transition is determined more by $L(p)$ which is a global property and remains obscured at the local level gauged by $C(p)$.

Additionally, Fig.~\ref{sw} sets out an interesting outcome: the probability range for which the small-world phenomena occurs is conditioned by the order of hyperedges. For a fixed rewiring probability, the larger the order of the hyperedges, the smaller the clustering coefficient of nodes owing to a larger number of neighbors and thus a larger number of possible hyperedges between them. Consequently, in the definition of the clustering coefficient, the denominator prevails over the numerator and favors the decrease in $C$. However, the average shortest path length defined as the number of hyperedges required to reach the target node from the source node remains independent from the order of hyperedges. Accordingly, small-world phenomena occur at lower $p$ for higher-order hyperedges.
The key result here is that for intermediate values of $p$, the hypergraph is a small-world hyper network, featuring prominent clustering attributes (like hyper regular lattices) and yet having  small average path lengths (as random hypergraphs). The result obtained is consistent with different orders of uniform hypergraphs considered here from $Q=2$ to $Q=6$.
\begin{figure}[t]
	\centering
	\includegraphics[width= 0.5\textwidth]{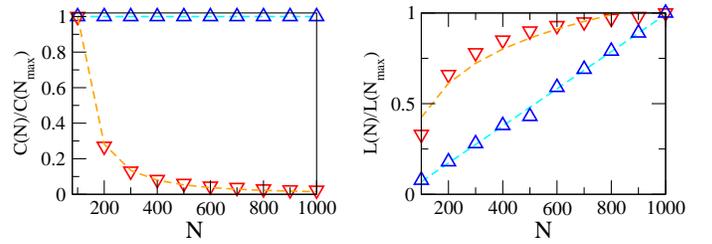}
	\caption{(Color online) Left: Clustering coefficient $C(N)$ and Right: average shortest path length $L(N)$ as a function of network size $N$ for two limiting cases, $3$-hyper regular lattice $p=0$ ($\color{blue} \triangle$) and $3$-hyper random graph $p=1$ ($\color{red} \triangledown$). Curve fitting of the data suggests that $C(N)_{p=0}$ is independent of the network size and $C(N)_{p=1}\propto \frac{1}{N}$,  whereas $L(N)_{p=0}\propto N$ and $L(N)_{p=1}\propto \ln{N}$. Each data point is averaged over $20$ random realizations.  }
	\label{size}
\end{figure}

\textit{Interplay with network measures:} Next, we discuss the correspondence between the structural properties ($C$ and $L$) and the network measures ($N$ and $k$) for $3$- uniform hypergraphs. To do so, we first calculate the neighbors of each node for the case of regular hypergraphs. The total number of neighbors can be calculated as follows.
The first hyperedge connecting node $i$ with its nearest neighbors (yellow colored in Fig.~\ref{sch2}) contributes $2$ neighbor nodes, the next hyperedge connecting next nearest neighbors (blue colored) contributes $1$ more neighbor node and then the hyperedge connecting next to next neighbor nodes contribute $1$ more and so on. In this way on each side (right and left) the number of  neighbors is $2+\frac{k}{2}-1$ and thus the total number of neighbors is $nb=k+2$. This can be generalized to $Q$- uniform hypergraphs as $Q-1$ neighbor nodes contributed by first hyperedge connecting nearest neighbors, then $1$ neighbor node by other following hyperedges, so the total number of neighbors is $nb=k+2Q-4$. For pairwise interaction networks, the latter quantity reduces to $nb=k$.

Now, $C_{i}$ is defined as the ratio of the actual number of edges between the neighbors of node $i$ to the possible number of edges between the neighbors. For $p=0$ one first calculates the actual number of hyperedges between $nb_i$ (neighbor nodes of $i_{th}$ node) as follows.
The nearest nodes (nodes $2$ and $12$ in Fig.~\ref{sch2}) contribute $k-2$ hyperedges each between the neighbors, then the next nearest neighbor nodes contribute $k-4$ hyperedges each (nodes $3$ and $11$), next to next nearest neighbor nodes contribute $k-4$ hyperedges each and the process goes on until $\frac{k}{2}$ nodes on each side. Thus, the numerator becomes $2((k-2)+(k-4)+(k-6)+\dots+(k-k))$ which is equal to $k(\frac{k}{2}-1)$, whereas the possible number of hyperedges between the neighbors can be calculated as $\binom{nb_i}{Q}$
$=\frac{k(k+1)(k+2)}{3!}$. Thus, for regular hypergraphs, the clustering coefficient of node $i$ becomes
\begin{equation}
C_i=\frac{3(k-2)}{(k+1)(k+2)}
\end{equation}
which tends to $\frac{1}{k}$ in the limit of large $k$. Next, for random hypergraphs with connection probability $\epsilon$ defined as the probability that a node $i$ belongs to a hyperedge $e$: $\epsilon=\frac{k}{\binom{N-1}{Q}} \sim \frac{k}{N^2}$, the clustering coefficient of node $i$ can be calculated as $C_i =\frac{\epsilon\binom{nb_i}{Q}}{\binom{nb_i}{Q}} \sim \frac{k}{N^2}$. 

Next, one calculates the average shortest path length for regular hypergraphs with average degree $k$ and $nb$ neighbors of each node as follows. In one step, one can reach $\frac{nb}{2}$ nodes in either direction from node $i$. Similarly, in $2$ steps, $2(\frac{nb}{2})$ nodes can be reached and then $3(\frac{nb}{2})$ nodes in $3$ steps and so on, until $\frac{N}{2}=L\frac{nb}{2}$\ which implies $L=\frac{N}{nb}$. In the case of $3$- uniform hypergraphs, one has
\begin{equation}
L=\frac{N}{k+2} \sim \frac{N}{k}
\end{equation}
in the limit of large $k$. For random networks, instead, node $i$ has $nb$ neighbors, these $nb$ neighbors also has $nb$ neighbors so node $i$ has $nb^2$ second neighbors. Similarly, $nb^3$ third neighbors, $nb^4$ fourth neighbors, and so on. In order to reach all $N$ nodes in the graph, one must have $nb^L=N$, which leads to $
L\sim\frac{\ln{(N)}}{\ln{(nb)}}$.

\begin{figure}[t]
	\centering
	\includegraphics[width= 0.5\textwidth]{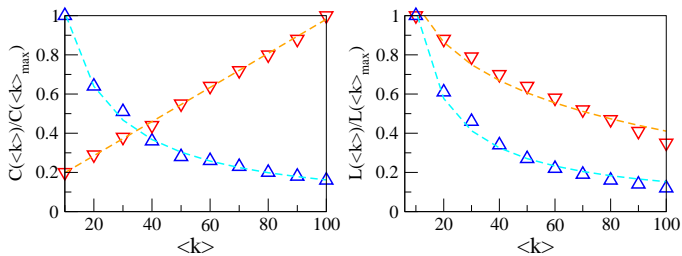}
	\caption{(Color online) Left: Clustering coefficient $C(N)$ and Right: average shortest path length $L(N)$ as a function of average degree $k$ for two limiting cases, $3$-hyper regular lattice $p=0$ ($\color{blue} \triangle$) and $3$-hyper random graph $p=1$ ($\color{red} \triangledown$). Curve fitting of the data suggests $C(k)_{p=0}\propto \frac{1}{k}$ and $C(k)_{p=1}\propto k$  whereas $L(k)_{p=0}\propto \frac{1}{k^2}$ and $L(N)_{p=1}\propto \ln{N}$. Each data point is averaged over $20$ random realizations.  }
	\label{degree}
\end{figure}

Also, we numerically calculate $C$ and $L$ as a function of network size ranging from $N=100$ to $N=1,000$, while keeping degree and hyperedge order fixed i.e., $Q=3$ (Fig.~\ref{size}). After curve fitting the obtained plots, one finds that, as p$\rightarrow0$, $C$ is independent of system size while $C_{random} \sim \frac{1}{N^2}$ (Fig.~\ref{size} (left)). However, $L_{p=0}$ increases linearly with system size whereas $L_{random} \sim \ln{N}$ (Fig.~\ref{size} (right)).

Similarly, one may compute $C$ and $L$ for different values of the average degree $k$, while keeping $N$ and $Q$ fixed (Fig.~\ref{degree}). After fitting, one finds that for $p\rightarrow0$ $C \sim \frac{1}{a+bk}$, with $a=0.4$ and $b=0.05$ whereas for random hypergraphs, $p=1$, $C$ increases linearly with $k$ (Fig.~\ref{degree} (left)). Also, $L_{p=0}$ is observed to vary as $\frac{1}{k}$ while $L_{random} \sim \frac{1}{ln(k)}$. These results suggest that hyper regular lattices are characterized by high clustering and large distances ($L \sim N$) while hyper random networks ($p=1$) are weakly clustered and feature the scaling $L \sim \ln{N}$. The above findings validate the theoretical calculations of $C$ and $L$ dependence on network measures ($N$ and $k$).

\textit{Conclusion:} Inspired by the Watts-Strogatz algorithm, we have considered $Q$-uniform hypergraphs and we have focused on a hyperedge rewiring method where hyper regular lattices are rewired with probability $p$. We first have calculated the  clustering coefficient $C$ and the average shortest path length $L$, and we have individuated, in all cases, a small-world transition. The nature of this transition is the same for different $Q$- uniform hypergraphs, but an increase in the hyperedge order shrinks the range of rewiring probabilities for which the transition occurs.
Additionally, we evaluated the relationship between structural properties ($C$ and $L$) and network measures for $3$- uniform hypergraphs. Our findings suggest that, as $p\rightarrow0$, $C\sim \frac{1}{k}$ and $L\sim\frac{N}{k}$ whereas for $p\rightarrow1$, $C\sim\frac{k}{N^2}$ and $L\sim\frac{\ln{N}}{\ln{k}}$. The average shortest path length for the  two limiting cases $p=0$ and $p=1$ seems to follow the same behavior as in the case of pairwise interactions. The present work can be extended to mixed hypergraphs where hyperedges of different order are present together in a hypergraph so that the functional significance of present results can be studied in real-world systems. Other directions for future studies could be the study of processes, for example, diffusion or contagion on small-world hypergraphs. 

\begin{acknowledgments}
We acknowledge VAJRA project VJR/2019/000034 under which this work was carried out. SJ gratefully acknowledges SERB Power grant SPF/2021/000136. The work is supported by the computational facility received from the Department of Science and Technology (DST), Government of India under FIST scheme (Grant No. SR/FST/PSI-225/2016). We thank Subhasanket Dutta for useful suggestions and interesting discussions.
\end{acknowledgments}


\begin{thebibliography}{99}

\bibitem{Milgram1967}  S. Milgram. The small world problem. Psychology Today, 2, 60-67 (1967).

\bibitem{Karinthy1967} F. Karinthy. Chains in "Everything is diﬀerent", Budapest, (1929).

\bibitem{PoolKocher1978} I. de S. Pool and M. Kochen. Contacts and inﬂuence, Social Networks, 1 (1978).

\bibitem{WattsStrogatz1998} D. Watts and S. Strogatz. Collective dynamics of ‘small-world’ networks. Nature 393, (1998).

\bibitem{internet} R. Albert, H. Jeong, A.L. Barabási. Diameter of the world-wide web. Nature 401, 130–131 (1999); A. Broder, R. Kumar, F. Maghoul, P. Raghavan, S. Rajagopalan, R. Stata, A. Tomkins, J. Wiener, Graph structure in the web, Computer Networks, 33, 1–6, (2000).

\bibitem{airt} L.A.N. Amaral, A. Scala, M. Barthélémy, and H.E. Stanley. Classes of small-world networks. PNAS 97, 21, (2000).

\bibitem{polymers} S.N. Jespersen, I.M. Sokolov, A. Blumen. Small-world rouse networks as models of cross-linked polymers. Journal of Chemical Physics, 113(17) 2000.

\bibitem{brain} D.S. Bassett, and E. Bullmore. Small-world brain networks. The Neuroscientist. 12(6), 2006.

\bibitem{metabolic} A. Wagner, and D.A. Fell. The small world inside large metabolic networks. Proceedings of the Royal Society of London. Series B: Biological Sciences, 268, (2000).

\bibitem{new} S. Alamdari. Small-world formation via local information. arXiv:2301.00849; M. Zhang, L. Li, G. Trajcevski, A. Zufle, and X. Zhou, Parallel hub labeling maintenance with high efficiency in dynamic small-world networks. IEEE Transactions on Knowledge and Data Engineering (2023); A.S. Reis, E.L. Brugnago, R.L. Viana, A.M. Batista, K.C. Iarosz, I.L. Caldas. Effects of feedback control in small-world neuronal networks interconnected according to a human connectivity map. Neurocomputing, 518, (2023).

\bibitem{hbrain} S. Gu, M. Yang, J.D. Medaglia, R.C. Gur, R.E. Gur, T.D. Satterthwaite and D.S. Bassett. Functional hypergraph uncovers novel covariant structures over neurodevelopment. Human brain mapping, 38, (2017).

\bibitem{hsocial}  V. Zlati, G. Ghoshal, and G. Caldarelli. Hypergraph topological quantities for tagged social networks. Physical Review E 80, 036118 (2009); J. Zhu, J. Zhu, S. Ghosh, W. Wu, and J. Yuan. Social influence maximization in hypergraph in social networks. IEEE Transactions on Network Science and Engineering 6, 801 (2018)

\bibitem{heco}  S.G. Alicia, B. Djordje, L. Osborne Melisa, J.F. Poyatos and S. Alvaro. High-order interactions distort the functional landscape of microbial consortia. PLoS Biology, 17, (2019).

\bibitem{hbio} S. Klamt, U.U. Haus, and F. Theis. Hypergraphs and cellular networks. PLoS computational biology 5, e1000385 (2009); S. Feng, E. Heath, B. Jefferson, C. Joslyn, H. Kvinge, H.D. Mitchell, B. Praggastis, A.J. Eisfeld, A.C. Sims, L.B. Thackray, et al. Hypergraph models of biological networks to identify genes critical to pathogenic viral response. BMC bioinformatics 22, 1 (2021).

\bibitem{hevol} U. Alvarez-Rodriguez, F. Battiston, G. F. de Arruda, Y. Moreno, M. Perc, and V. Latora. Evolutionary dynamics of higher-order interactions in social networks. Nature Human Behaviour 5, 586 (2021); G. Burgio, J. T. Matamalas, S. Gomez, and A. Arenas. Evolution of cooperation in the presence of higher-order interactions: From networks to hypergraphs. Entropy 22, 744 (2020).

 \bibitem{hpro} T. Gaudelet, N. Malod-Dognin and N. Prˇzulj. Higher-order molecular organization as a source of biological function. Bioinformatics, 34, (2018); N. Franzese, A. Groce, T. Murali, and A. Ritz. Hypergraph-based connectivity measures for signaling pathway topologies. PLoS computational biology, 15, (2019).

 \bibitem{signal} A. Ritz, A.N. Tegge, H. Kim, C.L. Poirel and T. Murali. Signaling hypergraphs. Trends in biotechnology, 32, 356–362 (2014).


\end{thebibliography}
\end{document}